\documentclass[lettersize,journal]{IEEEtran}
\usepackage[T1]{fontenc}
\usepackage[utf8]{inputenc}
\usepackage{amsmath,amsfonts}
\usepackage{array}
\usepackage{textcomp}
\usepackage{stfloats}
\usepackage{url}
\usepackage{graphicx}
\usepackage{cite}
\usepackage{comment}

\newif\ifanonymous
\anonymousfalse

\begin{document}

\title{Exploring the Effects of Personality in Human-Agent Interactions: A Study on User-Agent Synchrony with Human-based Vocalics}

\ifanonymous
\author{Anonymous Author(s)}
\else
\author{Kai Alexander Hackney, Jhonathan Sora-Cardenas, Aibek Musaev, and Pedro Guillermo Feijóo-García,~\IEEEmembership{Member,~IEEE}%
\thanks{The authors are with the College of Computing of the Georgia Institute of Technology, Atlanta, Georgia, USA.}%
\thanks{Corresponding author: Pedro Guillermo Feijóo-García (e-mail: pfeijoogarcia@gatech.edu).}%
}
\fi

\markboth{IEEE Transactions on Affective Computing}%
{\ifanonymous Anonymous Author(s)\else Hackney et al.\fi}

\maketitle

\begin{abstract}

    User trust is paramount in human-agent interactions, as it allows users to feel comfortable being themselves around an agent. The process of building user rapport starts in how an agent was designed, from its modality to the setting, to any of the many features and characteristics that can be tailored. All these aspects can affect whether users will be able to properly interact with the virtual agent and achieve the intended purpose. One such feature that is critical in human-human interactions is personality. A person's personality can strongly influence whether those that interact with them perceive them as trustworthy. This study used virtual agents generated from human voices with known personality traits to evaluate user perceptions. We found that extraverted agents were deemed to be more likeable by users, and that there was no significant effect of user-agent synchrony on user perceptions of the agent. In addition, it was found that user personality, without regard for agent personality, affected user perceptions of the agents. Our observations suggest that user perceptions may depend more on agent-topic synchrony than user-agent synchrony, and contribute to the broader community with considerations for the design of effective human-agent interactions.

\end{abstract}


\section{Introduction}

 In human-human interactions, perceptions of trustworthiness and credibility are influenced by a range of factors, such as prior interactions, accumulated experiences, background knowledge, and personality traits \cite{fiske2007universal}. When designing virtual humans (e.g., embodied conversational agents), past interactions are difficult to control for and are less salient as an effective agent is expected to perform well from the very first interaction. While an agent's expertise can be controlled through access to relevant information, personality stands out as a factor that may significantly influence user perceptions, mirroring its importance in human-human interactions \cite{10.1145/3514197.3549686}.

 The Computers Are Social Actors (CASA) theory shows that users interact with virtual agents in a similar manner to how they interact with humans \cite{10.1145/3386867, 10.1037/1076-898x.7.3.171}. In human interactions, attributes such as facial features, clothing, gait, overall appearance, and voice are closely related to personality \cite{borkenau1995observable}. Virtual agents replicate this through their design modalities, including facial features \cite{ghosh2023evaluating}, overall appearance \cite{ghosh2023evaluating}, clothing \cite{parmar2018looking}, gait \cite{gait2018perception}, and voice \cite{10.1145/3386867}. Among these modalities, voice is distinctive in that, even when other channels are present, it has been shown to convey rich social information on its own, including personality traits, emotional state, and trustworthiness \cite{maltezou2025how, haresamudram2024talking}.

 A growing number of deployed agents are designed to be voice-only (including smart speakers, interactive voice response (IVR) systems, and telephone-based assistants), making voice a central modality in everyday human-agent interaction \cite{10.1145/3386867}. Reflecting this, voice in agents has been studied from multiple angles, such as perceived trust in voice-enabled smart home systems \cite{liu2021}. In particular, the link between voice and perceived personality is relevant for the present work, as listeners form distinct personality impressions directly from agent voices \cite{kim2025}.
 
  Within this line of work, one phenomenon that has been explored quite a bit is the similarity-attraction effect, which says that humans are predisposed to feel positively about those who have similar personalities \cite{BYRNE196935}. A growing body of research has explored whether the similarity-attraction effect appears in human-agent interactions, specifically with extraversion as a factor being explored \cite{10.1145/3570945.3607318}. For instance, Nass \& Lee \cite{10.1037/1076-898x.7.3.171} observed that participants preferred agents with personalities that aligned with their own. However, a major limitation of prior work lies in its reliance on synthesized voices generated from calculated parameters, which may fail to capture the nuance of the corresponding human personalities \cite{app11188776, 10.1037/1076-898x.7.3.171}. Existing research highlights the need for studies using user-based voices to better assess whether the similarity-attraction effect holds in more authentic settings. Such work would be valuable across a range of contexts, including mental health, medical, and other social interactions \cite{gratch2014s, 10.1145/3472306.3478367}. 

 This study builds upon prior work that examined whether users can perceive personality traits in AI-generated voices when those voices are based on real human recordings \cite{hackney2026exploringinterplayvoicepersonality}. In that study, participants first completed a Ten Item Personality Inventory (TIPI) \cite{TIPI} and recorded their voices. Based on the results, participants who ranked highest in introversion and extraversion were selected to generate four virtual agent voices: male introvert, male extravert, female introvert, and female extravert. A separate group of participants then listened to these voices and evaluated their perceived personalities using the same TIPI \cite{10.1145/3570945.3607318}. The findings indicated that participants could reliably perceive differences in extraversion among the female agents, while being unable to significantly differentiate the male agents along this dimension. Additionally, some users demonstrated a tendency to perceive agents as having personality traits similar to their own.

 Building on these findings, the present study leverages the same set of agents to examine the role of user-agent synchrony in shaping perceptions of agents \cite{10.1145/3290605.3300270, 10.1145/3232077}. Participants first completed a TIPI \cite{TIPI} assessment to establish their personality profile. They then listened to two agents, either both male or both female, presented in randomized order, and evaluated the perception the participant had of each agent using a standardized instrument. This mixed design (within and between subjects) allowed us to analyze the relationship between personality similarity and perceived credibility, providing insight into whether users were more inclined to perceive agents that reflected their own personality traits in a positive manner.

 By examining whether the similarity-attraction effect extends to AI-generated voices grounded in real human data, this study aimed to bridge a gap in existing research and inform the design of more credible and effective virtual agents \cite{10.1145/3386867, app11188776}.

 This study addressed the following research question: 

\begin{itemize}
    \item \textbf{RQ1}: To what extent does the similarity-attraction effect remain when AI-generated voices are derived from real human voices with known personalities, rather than artificially constructed ones?
\end{itemize}


\section{Background}

\subsection{CASA and the Similarity-Attraction Effect in Human-Agent Interaction}

One of the most studied phenomena in human-agent interactions is the CASA theory, which states that human-computer interactions widely mirror those of human-human interactions. Existing research has attempted to explore this theory with respect to various modalities of computers \cite{10.1145/3232077}. Bernier \& Scassellati (2010) found that CASA theory extends to human interactions with embodied robotics, showing that humans are capable of anthropomorphizing wide varieties of non-human agents \cite{5578828}.

Within human-human interactions, a phenomenon that ties directly into perceptions of credibility is the similarity-attraction effect. The similarity-attraction effect suggests that people are more likely to develop positive feelings, trust, and credibility with agents that resemble them in some way \cite{BYRNE196935, gratch2014s}. As CASA has been explored in more depth, previous research appeared to show that the similarity-attraction effect has a significant effect on user perceptions of non-human agents \cite{5578828}.
The influence of similarity-attraction is frequently moderated by situational context. For instance, Maxim et al. \cite{10.1145/3570945.3607318} observed a generalized preference for introverted agents in mental health settings, suggesting that users may feel more comfortable being in control of the conversation in sensitive environments. Similarly, the dominance or submissiveness of an agent can alter its appeal depending on the user's specific role \cite{10.1145/3687272.3690913}. These findings suggest that cohesion and continuity, the alignment between an agent's appearance, voice, and context, may be more critical to user perception than simple similarity \cite{10.1145/3386867}.

\subsection{Complementary vs. Similarity Attraction in User-Agent Synchrony}

The Computers Are Social Actors (CASA) paradigm remains a foundational framework for evaluating human-agent interaction, yet the reliability of translating human-human social phenomena to virtual environments remains a subject of active debate \cite{9377323}. Early research prioritized complementary attraction: the preference for agents with opposing traits. Despite this, recent evidence suggests that similarity-attraction may be more pervasive across non-human interactions. Bernier \& Scassellati \cite{5578828} demonstrated that rapport often hinges on situational contexts, with users favoring agents that mirror their speech content, personality, or hobbies. This is reinforced by Kim et al. \cite{9377323}, who found that high perceived agreeableness in agents fosters trust even in the absence of explicit similarity. Thus, user preferences appear split, as similarity is preferred for emotional support, but authoritative or knowledgeable agents are often favored for task-oriented decision-making.

\subsection{Voice and Perceived Personality}

A growing number of deployed agents are voice-only by design (including smart speakers, interactive voice response (IVR) systems, and telephone-based assistants) making voice a central modality in everyday human-agent interaction \cite{10.1145/3386867, liu2021}. A synthesis of this growing body of work and show that voice in agents conveys a wide range of psychosocial information, including personality, anthropomorphism, gender, age, and affect, as well as sociolinguistic features such as paralanguage, accent, and style of speech \cite{10.1145/3386867}. Even when other modalities are present, voice may play a particularly strong role: comparison of voice-only smart speaker with an embodied robot across four levels of voice humanness and found that, although body and voice each contributed independently to anthropomorphism, voice exerted a stronger influence than physical appearance on agent perception (excluding trust, where no significant effects emerged)\cite{haresamudram2024talking}.

Within this body of work, personality has emerged as a particularly salient dimension transmitted through voice. A critical factor of user perception is the alignment between agent personality and vocal characteristics. Research indicates that traits such as introversion and extraversion significantly influence user preference, which further validates the similarity-attraction effect \cite{10.1145/3570945.3607318, 10.1145/3290605.3300270}. Vocal parameters, including pitch, tone, and speech rate, directly shape perceptions of competence and reliability \cite{10.1145/3386867}. Classic work in social psychology has further shown that, among personality traits, extraversion is the trait most reliably inferred from voice quality, with vocal cues such as vocal effort, dynamic range, loudness, and sharpness serving as ecologically valid indicators \cite{Scherer1978}. More recently, Kim et al. \cite{kim2025} had listeners rate the perceived personality of natural and synthetic voices using the Big Five model and extracted acoustic parameters including voice quality, fundamental frequency, and articulation rate; they found that acoustic differences between synthetic voices and their human counterparts produce distinct effects on personality perception, indicating that even fine-grained acoustic features shape how listeners attribute personality to a voice. Beyond static vocal traits, vocal synchrony also plays a role: Nishimura et al. \cite{app11062502} demonstrated that adapting an agent's vocal cadence to match the user's speech patterns in real time significantly enhances rapport and engagement.

\subsection{Limitations of Synthesized Voices and Voice-Based Personality Research}
Despite this growing body of work, the reliance of the existing literature on synthesized voices presents a methodological limitation: these voices often lack the prosodic nuances of natural human speech, which could lead to results based on stereotypical exaggerations rather than authentic personality traits \cite{app11188776}. If AI agents fail to mirror the subtleties of real human personalities, current findings may not accurately reflect real-world user preferences. Existing research shows that users prefer agents with similar traits, but it remains unclear whether this preference holds when agent voices are derived from real human recordings rather than constructed from calculated parameters. This study addresses this gap by replicating personality alignment research using virtual agents generated from real human voices with known personality profiles.

\section{Methods}

This user study took place across the Summer and Fall of 2025 within a variety of courses from a Southeastern University in the United States (USA). The study was approved by the Institutional Review Board (IRB) at the Georgia Institute of Technology, and data were collected asynchronously over one-week periods in Summer and Fall 2025 across multiple courses, taking place through Qualtrics \cite{qualtrics}.

\subsection{Participants}
A total of 189 adult (18 years of age or older) participants were recruited from the institution. Recruitment was conducted regardless of gender, age, ethnicity, or place of origin. Of our participants, only 146 completed the survey and provided informed consent; the responses of any participants who failed to complete the survey or provide informed consent will not be used. Participants were enrolled in the Fall/Summer of 2025 and were recruited from multiple courses within the College of Computing. Participants reported ages were 18 years of age (n = 15), 19 years of age (n = 56), 20 years of age (n = 41), 21 years of age (n = 18), 22 years of age (n = 7), and 23-29 years of age (n = 9). Participants reported their spoken languages; 70 were monolingual and 76 were multilingual. Participants reported being from 24 different countries: the USA (n = 101) or other places (n = 45). Participants’ reported gender distribution was male (n = 106), female (n = 34), or other (n = 4). Two participants chose not to identify with any gender (n = 2).

\subsection{Procedure}
This study was conducted online and asynchronously via Qualtrics \cite{qualtrics}. This was done to make participation accessible, allowing for more participants to respond. The study included both within-subjects and between-subjects comparisons, enabling direct comparisons between agents with different personality traits. The median time taken on the survey was 20 minutes, matching the expectations outlined in our informed consent. 

Participants were evenly and randomly assigned to one of four groups, where the distinguishing feature of each group was the agents listened to and their order. The four groups were:

\begin{enumerate}
    \item Listening to the extraverted male agent, then listening to the introverted male agent. Labeled as "ME->MI" in Figure \ref{fig:Male Godspeed T-Test}.
    \item Listening to the introverted male agent, then listening to the extraverted male agent. Labeled as "MI->ME" in Figure \ref{fig:Male Godspeed T-Test}.
    \item Listening to the extraverted female agent, then listening to the introverted female agent. Labeled as "FE->FI" in Figure \ref{fig:Female Godspeed T-Test}.
    \item Listening to the introverted female agent, then listening to the extraverted female agent. Labeled as "FI->FE" in Figure \ref{fig:Female Godspeed T-Test}.
\end{enumerate}

The four agents mentioned above consisted of a voice recording generated using ElevenLabs \cite{elevenlabs} based on a voice sample taken from \cite{hackney2026exploringinterplayvoicepersonality}. 

Eleven Multilingual V2 was the model used to generate all four voices (with settings of Stability: 50\%, Similarity: 75\%, Style Exaggeration: 0\%, and Speaker boost: on). To ensure that the output from ElevenLabs would be high-quality, the voice samples input into ElevenLabs \cite{elevenlabs} were comprehensive, including both the \textit{Grandfather} passage and the \textit{Rainbow} passage, two passages notable for capturing the full range of a person’s voice. The output script used was a passage taken from a Journal publication on job burnout in the Annual Review of Psychology \cite{annurev:/content/journals/10.1146/annurev.psych.52.1.397}. 

The voice samples to use were decided based on the participants’ responses to the TIPI--an inventory used to determine scores on each of the Big Five personality traits \cite{TIPI}. The two statements regarding extraversion within the TIPI include: \textit{``I see myself as Extraverted, enthusiastic"} and \textit{``I see myself as Reserved, quiet"}.  The users with the highest and lowest Extraversion scores for both male and female participants were identified. Their recorded voice samples were then assessed for quality and utilized to create the agents. 

Extraversion was selected to differentiate personality, as prior research identifies it as both highly perceptible and commonly used in human-agent interaction studies \cite{10.1145/3570945.3607318, 10.1037/1076-898x.7.3.171}. Participants were only asked to listen to two of the agents to ensure that they could reasonably listen to both voices and answer questions about them in a single sitting. The study consisted of five main blocks.

\begin{figure}[!t]
 \centering
 \includegraphics[width=\linewidth]{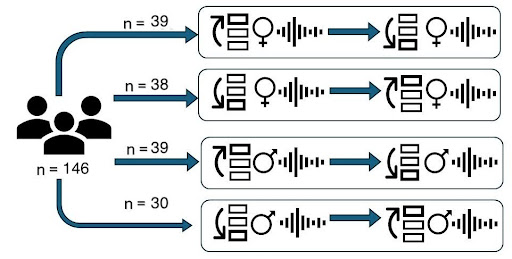}
 \caption{Sample Population and Participant Grouping}
 \label{fig:Protocol}
\end{figure}

\textbf{Block 1 - Informed consent} provided participants with Informed Consent information containing the details of the study and two multiple-choice questions. Participants were asked to read and respond (1) if they were 18 years of age or older and fluent in English and (2) if they agreed to participate.

\textbf{Block 2 - Participant Personality} asked participants to respond to the personality diagnostic (i.e., TIPI \cite{TIPI}). This block included ten statements that participants would rank on a scale of one to seven how much they agreed with (one being completely disagree and seven being completely agree). The TIPI was selected due to its brevity, which minimized participant burden and lowered the barrier to entry for data collection. Furthermore, utilizing this scale aligned directly with the foundational study used to design the agents, ensuring methodological consistency across measures.

\textbf{Block 3a - First Interaction} Participant listened to the first artificial voice based on their assigned group. After listening to the ~five-minute recording, the participant was then asked to respond, regarding the voice they just listened to, to a Godspeed Questionnaire \cite{godspeed}. User perceptions of credibility are difficult to isolate, so instead we relied upon one of the most cited questionnaires in human-robot interactions due to its comprehensive nature, measuring: anthropomorphism, animacy, likeability, perceived intelligence, and perceived safety. Through the questionnaire we gauged the complete perception that the user has of the agent. The Godspeed Questionnaire provides participants with two words at a time and asks the user to provide a number one through five (one being most like the first word and five being most like the second word). Some word pairings include: Dislike/Like, Unfriendly/Friendly, and Calm/Agitated. This use of the Godspeed Questionnaire to evaluate user perception of voice-based agents is consistent with prior work in which voice is manipulated as an experimental variable \cite{haresamudram2024talking, alimardani2024}.

\textbf{Block 3b - Second Interaction} Participant listened to the second artificial voice based on their assigned group. The participant then repeated answering the Godspeed Questionnaire \cite{godspeed} again for the second voice.

\textbf{Block 4 - Participants’ Demographics} contained six questions, asking about participant age, ethnicity, gender, country of origin, city of origin, and frequently used languages.

\textbf{Block 5 - Participants’ Contact Information} asked for email addresses to ensure compensation could be administered according to IRB protocol.

The data collection architecture was adapted from the paradigm established by Bernier \& Scassellati (2010) \cite{5578828}. Participants first completed a baseline assessment, interacted with an agent that was either behaviorally similar or dissimilar to their own profile, and subsequently evaluated their perceptions of the agent. However, whereas Bernier \& Scassellati focused on toy preferences, this study examines personality traits. Furthermore, we extended this baseline-interaction framework to encompass multiple sequential interactions, allowing for a more dynamic assessment over time.

\subsection{Data Analysis}
Participants were first analyzed concerning their responses to Block 2 and the TIPI Questionnaire within. From their responses to the TIPI Questionnaire, we were able to calculate an Extraversion score for each participant. We then analyzed their responses to the Godspeed Questionnaire \cite{godspeed} contained within Blocks 3a and 3b and calculated the scores for the first and second voice the participant listened to, respectively. Participant demographics were also considered, particularly gender. A variety of statistical tests were used to search for significant findings, specifically correlation, regression, descriptive statistics, and Mann-Whitney U tests.

\section{Results}

\subsection{Findings: Extraverted Agents are more Likeable}

Our analysis first evaluated the discernibility of intended personality traits across the virtual agents utilized in this study. Prior research indicates that users can successfully differentiate between extraverted and introverted female agents, though this distinction is often not statistically significant for male agent profiles \cite{hackney2026exploringinterplayvoicepersonality}. In order to properly analyze participant perceptions of agents without risk of contamination from their previous interaction, this analysis only utilized each participant's perception of the agent they first interacted with. For this interaction, all four voice profiles (male and female) were utilized across the experimental groups $(n = 146)$. 

When examining perceptions of female agents $(n = 77)$ through the Godspeed Questionnaire \cite{godspeed}, participants found the extraverted agent significantly more likeable (Mann-Whitney $U$ test, $p < .01$) and animated (Mann-Whitney $U$ test, $p = 0.013$) (see Table \ref{fig:Female Godspeed T-Test}). Conversely, no significant differences were observed regarding male agent profiles $(n = 69)$ (see Table \ref{fig:Male Godspeed T-Test}). These observations suggest that in contexts where vocal cues permit successful personality differentiation, extraversion is positively correlated with perceptions of animacy and likeability.


\begin{table*}[!t]
    \caption{User Godspeed Mann-Whitney U test comparison female extraverted and introverted Agents}
    \label{fig:Female Godspeed T-Test}
    \centering
    \begin{tabular}{|c|c|c|}
        \hline
        Mann-Whitney U test & U & p \\
        \hline
        Anthropomorphism & 778.0 & 0.709 \\
        \hline
        Animacy & 984.0 & 0.013 \\
        \hline
        Likeability & 997.5 & 0.008 \\
        \hline
        Intelligence & 854.0 & 0.249 \\
        \hline
        Safety & 697.5 & 0.657 \\
        \hline
    \end{tabular}
    \par\medskip
    \small
    \resizebox{\textwidth}{!}{%
\begin{tabular}{|c|c|c|c|c|c|c|c|c|}
        \hline
        Descriptives & Group & N & Mean & SD & SE & Coef. of Var. & Mean Rank & Sum Rank \\
        \hline
        Anthropomorphism & FE->FI & 39 & 2.903 & 0.920 & 0.147 & 0.317 & 39.949 & 1558.0 \\
        \hline
         & FI->FE & 38 & 2.795 & 0.936 & 0.152 & 0.335 & 38.026 & 1445.0 \\
        \hline
        Animacy & FE->FI & 39 & 2.867 & 0.882 & 0.141 & 0.308 & 45.231 & 1764.0 \\
        \hline
        & FI->FE & 38 & 2.316 & 0.940 & 0.152 & 0.406 & 32.605 & 1239.0 \\
        \hline
        Likeability & FE->FI & 39 & 3.349 & 0.838 & 0.134 & 0.250 & 45.577 & 1777.5 \\
        \hline
        & FI->FE & 38 & 2.789 & 0.899 & 0.146 & 0.322 & 32.250 & 1225.5 \\
        \hline
        Intelligence & FE->FI & 39 & 3.631 & 0.720 & 0.115 & 0.198 & 41.897 & 1634.0 \\
        \hline
        & FI->FE & 38 & 3.405 & 0.823 & 0.134 & 0.242 & 36.026 & 1369.0 \\
        \hline
        Safety & FE->FI & 39 & 3.487 & 0.717 & 0.115 & 0.206 & 37.885 & 1477.5 \\
        \hline
        & FI->FE & 38 & 3.561 & 0.813 & 0.132 & 0.228 & 40.145 & 1525.5 \\
        \hline
    \end{tabular}%
}
\end{table*}



\begin{table*}[!t]
    \caption{User Godspeed Mann-Whitney U test comparison male extraverted and introverted Agents}
    \label{fig:Male Godspeed T-Test}
    \centering
    \begin{tabular}{|c|c|c|}
        \hline
        Mann-Whitney U test & U & p \\
        \hline
        Anthropomorphism & 696.0 & 0.179 \\
        \hline
        Animacy & 621.5 & 0.661 \\
        \hline
        Likeability & 684.0 & 0.229 \\
        \hline
        Intelligence & 605.0 & 0.811 \\
        \hline
        Safety & 501.5 & 0.308 \\
        \hline
    \end{tabular}
    \par\medskip
    \small
    \resizebox{\textwidth}{!}{%
\begin{tabular}{|c|c|c|c|c|c|c|c|c|}
        \hline
        Descriptives & Group & N & Mean & SD & SE & Coef. of Var. & Mean Rank & Sum Rank \\
        \hline
        Anthropomorphism & ME->MI & 39 & 2.821 & 0.810 & 0.130 & 0.287 & 37.846 & 1476.0 \\
        \hline
         & MI->ME & 30 & 2.593 & 0.672 & 0.123 & 0.259 & 31.300 & 939.0 \\
        \hline
        Animacy & ME->MI & 39 & 2.441 & 0.721 & 0.116 & 0.296 & 35.936 & 1401.5 \\
        \hline
        & MI->ME & 30 & 2.400 & 0.666 & 0.122 & 0.278 & 33.783 & 1013.5 \\
        \hline
        Likeability & ME->MI & 39 & 3.113 & 0.635 & 0.102 & 0.204 & 37.538 & 1464.0 \\
        \hline
        & MI->ME & 30 & 2.993 & 0.766 & 0.140 & 0.256 & 31.700 & 951.0 \\
        \hline
        Intelligence & ME->MI & 39 & 3.564 & 0.554 & 0.089 & 0.155 & 35.513 & 1385.0 \\
        \hline
        & MI->ME & 30 & 3.593 & 0.477 & 0.087 & 0.133 & 34.333 & 1030.0 \\
        \hline
        Safety & ME->MI & 39 & 3.487 & 0.670 & 0.107 & 0.192 & 32.859 & 1281.5 \\
        \hline
        & MI->ME & 30 & 3.656 & 0.697 & 0.127 & 0.191 & 37.783 & 1133.5 \\
        \hline
    \end{tabular}%
}
\end{table*}


\subsection{Findings: Extraversion and User-Agent Synchrony}
We evaluated the relationship between user-reported personality traits (TIPI \cite{TIPI}) and perceived agent likeability (Godspeed Questionnaire \cite{godspeed}). We assessed the following hypothesis:

\textbf{H-“Extraversion Synchrony on Agent Likeability”:} The level of likeability a user attributes to a virtual agent is significantly influenced by the alignment (synchrony) between the user’s self-reported extraversion and the agent's vocal personality.

The hypothesis was initially rejected, as no significant overall correlation was observed between user extraversion and agent likeability with the first agent each participant interacted with (Spearman’s Correlations, $n = 146, \rho = 0.120, p = 0.149$). This lack of significance persisted when segmenting the analysis exclusively to extraverted agents $(n = 78, \rho = -0.021, p = 0.856)$. However, a detailed analysis of the introverted agent subgroup $(n = 68)$ revealed a significant positive correlation (Spearman’s Correlations, $\rho = 0.283, p = 0.019$). These findings suggest that higher user extraversion is associated with increased likeability for introverted agents, indicating a preference for user-agent dissimilarity in this specific context.




\subsection{Findings: Personality Affected Agent Views}
While extraversion did not yield a universal synchrony effect, several secondary personality traits $(n = 146)$ demonstrated seemingly significant positive correlations with Godspeed Questionnaire metrics \cite{godspeed} when analyzing each participant's perception of the first agent they interacted with. As illustrated in Table \ref{fig:TIPI & Godspeed Correlation}. As this was an exploratory analysis we utilized an initial cutoff of $p < .05$:

\begin{itemize}
    \item \textbf{Agreeableness} and 
    \begin{itemize}
        \item Animacy ($\rho = 0.173, p = 0.036$)
        \item Likeability ($\rho = 0.235, p = 0.004$)
        \item Intelligence ($\rho = 0.269, p = 0.001$)
    \end{itemize}
    \item \textbf{Conscientiousness} and Animacy ($\rho = 0.174, p = 0.036$).
    \item \textbf{Emotional Stability} and Likeability ($\rho = 0.188, p = 0.023$).
\end{itemize}

It is understood that when analyzing correlations with multiple tests that the possibility of Type 1 error false positives when reporting significance if correction is not used to revaluate the cutoff value. In order to address this Benjamini-Hochberg procedure \cite{https://doi.org/10.1111/j.2517-6161.1995.tb02031.x} was utilized with a formula of $\frac{k}{m} \cdot \alpha$ where $k$ is the rank of the p-value sorted from lowest to highest, $m$ is the number of tests, and $\alpha$ is the cutoff false discovery rate. This gives the following cutoff values: $0.002, 0.004, 0.008$. This means that only 2 of our 5 previously identified correlations remain significant: Agreeableness and Intelligence ($\rho = 0.269, p = 0.001$) as well as Agreeableness and Likeability ($\rho = 0.235, p = 0.004$).

These results indicate that participants who self-reported higher scores on certain personality traits considered by TIPI \cite{TIPI})-- such as Agreeableness; consistently reported more positive perceptions towards the virtual agents (Godspeed Questionnaire \cite{godspeed}). This effect may be due to known response bias of highly agreeable individuals to report others highly on socially desirable responding scales \cite{wetzel2016response}.


\begin{table*}[!t]
    \caption{Correlation between TIPI results and Godspeed results}
    \label{fig:TIPI & Godspeed Correlation}
    \centering
    \centering
    \small
    \resizebox{\textwidth}{!}{%
\begin{tabular}{|c|c|c|c|c|c|c|}
        \hline
        Variable & Spearman's Correlations & Extraversion & Agreeableness & Conscientiousness & Emotional Stability & Openness to Experiences \\
        \hline
        Anthropomorphism & Spearman's rho & 0.083 & 0.057 & 0.130 & 0.046 & 0.076 \\
        \hline
        & p-value & 0.317 & 0.497 & 0.118 & 0.585 & 0.362 \\
        \hline
        Animacy & Spearman's rho & 0.126 & $0.173^*$ & $0.174^*$ & 0.088 & 0.054 \\
        \hline
        & p-value & 0.130 & 0.036 & 0.036 & 0.290 & 0.517 \\
        \hline
        Likeability & Spearman's rho & 0.120 & $0.235^{**}$ & 0.070 & $0.188^*$ & 0.124 \\
        \hline
        & p-value & 0.149 & 0.004 & 0.398 & 0.023 & 0.136 \\
        \hline
        Intelligence & Spearman's rho & -0.081 & $0.269^{**}$ & 0.085 & 0.023 & 0.051 \\
        \hline
        & p-value & 0.334 & 0.001 & 0.306 & 0.787 & 0.541 \\
        \hline
        Safety & Spearman's rho & -0.053 & 0.117 & 0.067 & 0.075 & -0.016 \\
        \hline
        & p-value & 0.529 & 0.160 & 0.418 & 0.370 & 0.848 \\
        \hline
    \end{tabular}%
}
    \smallskip
    \footnotesize $^* p<.05, ^{**} p<.01$
\end{table*}

\section{Discussion}

Responding to the Research Question: \textit{``To what extent does the similarity-attraction effect remain when AI-generated voices are made using real human voices with known personalities, rather than artificially constructed ones?”}, our observations indicate that the similarity-attraction effect was largely absent regarding vocal extraversion when utilizing human-derived voice samples. While no significant overall relationship was found between user extraversion and agent likeability, the emergence of a significant correlation specifically within the introverted agent group suggests that likeability may be moderated by user-agent extraversion dissimilarity rather than mirroring effects. This divergence from traditional CASA \cite{9377323} literature aligns with findings by Maxim et al., who similarly reported limited evidence for similarity-attraction in specific agent-mediated contexts \cite{10.1145/3386867, 10.1145/3570945.3607318}.

Although prior findings established that participants can discern vocal extraversion for the agents we used for the current user study \cite{hackney2026exploringinterplayvoicepersonality}, the lack of user-agent synchrony suggests that the specific vocal profiles or the "Job Burnout" script did not elicit the expected social mirroring-- this is consistent with findings by Kim et al. \cite{9377323}. Furthermore, the general preference for extraverted agents indicates that topic-agent synchrony, the alignment between an agent’s vocal personality and the nature of the information delivered, may have had a stronger impact on the user perceptions of the agents in this study, similar to the results found by Maxim et al. \cite{10.1145/3570945.3607318}.

Several methodological constraints may have influenced these observations. Limitations include those previously identified in asynchronous, online study designs, such as experimental variability introduced by participants completing the interaction in multiple sittings rather than a controlled environment \cite{hackney2026exploringinterplayvoicepersonality}. Additionally, the length and nature of the encounter, a five-minute monologue, may have reduced social presence and caused user focus to diminish. As noted in prior literature, a two-way, adaptive interaction where agents adjust their vocal characteristics in real-time may be more effective in fostering rapport and trust.

Future work should prioritize exploring the interplay between agent personality and varying situational contexts, such as mental health support or professional coaching, as suggested by Maxim et al. \cite{10.1145/3570945.3607318}. Research should also expand to include other personality dimensions beyond extraversion and investigate how personality characteristics interact with demographic variables like gender and age to shape user acceptance.

Ultimately, these findings confirm that agent personality can impact user perception, though not necessarily through traditional synchrony. As virtual agents begin to occupy larger roles in professional and sensitive capacities, it is critical that designers and developers account for the nuances of vocal personality to optimize agent effectiveness and user trust.

\section{Conclusions}

This paper discussed the impact of vocal personality synchrony on human-agent interactions, specifically evaluating how the similarity-attraction effect manifests when agents utilize real human voice samples rather than traditional synthetic cues. The potential for AI-based systems to mirror user personalities, either through pre-interaction diagnostics or dynamic real-time adaptation, remains a strategic consideration for enhancing user trust and engagement \cite{BYRNE196935, 5578828, app11062502}.

Our observations suggest that while both agent and user personalities independently influence perceived likeability and animacy, they do not exhibit the significant interplay required to demonstrate a similarity-attraction effect within the specific dimensions of introversion and extraversion. Consequently, our findings indicate that while users still perceive virtual humans as social actors, the expected synchrony may be dampened by situational factors or role expectations \cite{10.1145/3570945.3607318}.

Responding to our research question, our results suggest that human-agent interaction phenomena observed with synthetic voices do not consistently generalize to naturalistic, human-derived vocalics. This identifies a critical gap in current scholarship, as the transition toward more nuanced vocal generation fundamentally alters user perception and expectations of the agent’s capabilities. The lack of a direct mirroring effect suggests that topic-agent synchrony, aligning an agent’s personality with its specific task, may be more essential for fostering credibility than simple personality alignment, particularly in sensitive contexts such as mental health support or education \cite{gratch2014s, 10.1145/3472306.3478367}.

In conclusion, as virtual agents technology becomes increasingly integrated into sensitive social roles, design strategies must move beyond analogous assumptions. These findings underscore the importance of considering the interplay between vocal modality, topic, and user personality to ensure the effectiveness and acceptance of virtual agents in real-world applications.

\section{Limitations}
Our study has some limitations that should be considered when interpreting its findings. First, participants experienced a brief and one-sided interaction with the virtual agent. While this design is consistent with prior research on human-agent interaction \cite{10.1145/3689822, 10.1145/3472306.3478367}, it could have limited opportunities for rapport building and other social processes through which similarity-attraction effects could be observed. Future iterations should examine these relationships in longer interaction setups, considering human-agent interaction in bidirectional conversations.

Similarly, all participants were exposed to the same script focused on burnout and workplace experiences. Although this topic was selected to provide a meaningful conversational context, the content itself could have influenced participant perceptions independently of the virtual agent characteristics under exploration. Future research should explore whether similar findings emerge across different conversational topics and domains. To support transparency and reproducibility, we provide the full script in the Appendix \ref{Appendix}.

Additional limitations arise from the ways personality was considered and measured in our study. Each personality condition was represented by a single speaker, making it difficult to fully disentangle the effects of the intended personality portrayals from speaker-specific vocal characteristics (e.g., pitch, timbre, accent). As a result, our findings should be interpreted as participant responses to holistic personality portrayals rather than isolated personality traits. Moreover, participant personality was measured using the TIPI \cite{TIPI}, which we selected to reduce participant burden and facilitate completion of the broader questionnaire. However, because each personality dimension is measured using only two items, the instrument could have posed lower reliability than more comprehensive personality assessments. Future work should employ multiple voice exemplars for each personality condition and replicate our analyses using more comprehensive personality inventories with stronger psychometric properties.

The composition of our participant sample should also be considered. Our sample was not gender-balanced, reflecting the demographics of the population from which participants were recruited. While this distribution is representative of the recruitment context, it could limit the generalizability of our findings. Future research should seek more balanced samples to further investigate how user demographics shape perceptions of agents' characteristics in human-agent interactions.

Finally, some findings emerged from exploratory subgroup analyses conducted after the primary analyses were completed. Particularly, the relationships observed for the introverted virtual agent condition were not specified as primary hypotheses prior to our data collection. These results should therefore be interpreted as exploratory and hypothesis-generating rather than confirmatory evidence. Future studies should be specifically designed and powered to evaluate these relationships in more controlled experimental setups, through a priori hypotheses and independent replication efforts. 

\section*{Acknowledgments}
The authors thank all individuals who voluntarily participated in this study. Their input was essential for the outcomes reported in this manuscript.

The authors acknowledge the use of Grammarly and Gemini 3 Thinking (developed by Google), and Claude (developed by Anthropic) for spell-checking, grammar, and editing assistance. Both tools were used to refine the language and enhance the manuscript's clarity. No content, figures, or images were generated by artificial intelligence (AI) for this work.

\ifanonymous\else
\section*{Acknowledgments}
The authors thank all individuals who voluntarily participated in this study. Their input was essential for the outcomes reported in this manuscript.

This material is based upon work supported by the National Science Foundation (NSF) under Grant No. 2434428. Any opinions, findings, or recommendations expressed in this material are those of the author(s) and do not necessarily reflect the views of the NSF.

The authors acknowledge the use of Grammarly and Gemini 3 Thinking (developed by Google), and Claude (developed by Anthropic) for spell-checking, grammar, and editing assistance. Both tools were used to refine the language and enhance the manuscript's clarity. No content, figures, or images were generated by artificial intelligence (AI) for this work.

\fi

\bibliographystyle{IEEEtran}
\bibliography{references}

\appendices


\section{Agent Script on Job Burnout}\label{Appendix}

The relationship that people have with their work, and the difficulties that can arise when that relationship goes awry, have been long recognized as a significant phenomenon of the modern age. The use of the term burnout for this phenomenon began to appear with some regularity in the 1970s in the USA, especially among people working in the human services. This popular usage was presaged by Greene's 1961 novel, A Burn-Out Case, in which a spiritually tormented and disillusioned architect quits his job and withdraws into the African jungle. Even earlier writing, both fictional and nonfictional, described similar phenomena, including extreme fatigue and the loss of idealism and passion for one's job. What is noteworthy is that the importance of burnout as a social problem was identified by both practitioners and social commentators long before it became a focus of systematic study by researchers.

The evocative power of the burnout term to capture the realities of people's experiences in the workplace is what has made it both important and controversial in the research field. As the “language of the people,” burnout was more grounded in the complexities of people's relationship to work and gave new attention to some aspects of it. However, burnout was also derided at first as nonscholarly “pop psychology.” Unlike other research on the workplace, which used a top-down approach derived from a scholarly theory, burnout research initially utilized a bottom-up or “grass-roots” approach derived from people's workplace experiences. At first, the popular, nonacademic origins of burnout were more of a liability than an advantage. However, given the subsequent development of theoretical models and numerous empirical studies, the issue of research scholarship has now been laid to rest.

What has emerged from all of this research is a conceptualization of job burnout as a psychological syndrome in response to chronic interpersonal stressors on the job. The three key dimensions of this response are an overwhelming exhaustion, feelings of cynicism and detachment from the job, and a sense of ineffectiveness and lack of accomplishment. The exhaustion component represents the basic individual stress dimension of burnout. It refers to feelings of being overextended and depleted of one's emotional and physical resources. The cynicism (or depersonalization) component represents the interpersonal context dimension of burnout. It refers to a negative, callous, or excessively detached response to various aspects of the job. The component of reduced efficacy or accomplishment represents the self-evaluation dimension of burnout. It refers to feelings of incompetence and a lack of achievement and productivity at work.

Burnout was initially a very slippery concept—there was no standard definition of it, although there was a wide variety of opinions about what it was and what could be done about it. Different people used the term to mean very different things, so there was not always a basis for constructive communication about the problem and solutions for it. However, there was actually an underlying consensus about three core dimensions of the burnout experience, and subsequent research on this issue led to the development of a multidimensional theory of burnout. This theoretical framework continues to be the predominant one in the burnout field.

This would all lead to the identification of five common and distinct elements that help identify burnout. Dysphoric symptoms such as mental or emotional exhaustion, emphasis on mental and behavioral symptoms over physical ones, decreased work performance exhibited from individuals who did not exhibit any signs prior to their degraded performance, often being work-related, and often resulting in negative attitudes are all clear indicators of burnout manifesting in an individual. Since burnout can be encountered in various distinct scenarios, it is important to take into account an individual's unique context. Understanding and deconstructing burnout phenomena is the current focus in research efforts. \cite{annurev:/content/journals/10.1146/annurev.psych.52.1.397}

\begin{IEEEbiography}[{\includegraphics[width=1in,height=1.5 in,clip,keepaspectratio]{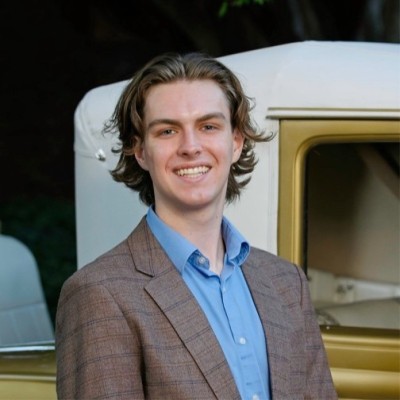}}]{Kai A. Hackney}
received the B.Sc. degree in computer science from the Georgia Institute of Technology, Atlanta, GA, USA, in 2026. He is currently pursuing the M.Sc. degree in computer science at the same institution, with a concentration in machine learning.

He is an active graduate researcher in the People-Agents Research for Computing Education (PARCE) Laboratory. He has conducted research under the supervision of Dr. Pedro Guillermo Feijóo-García during both his undergraduate and current graduate studies. His undergraduate concentrations were intelligence and modeling and simulation. He has also served as a Teaching Assistant for courses in machine learning, introductory artificial intelligence, and object-oriented programming.
\end{IEEEbiography}

\vspace{-1em}
\begin{IEEEbiography}[{\includegraphics[width=1in,height=1.5in,clip,keepaspectratio]{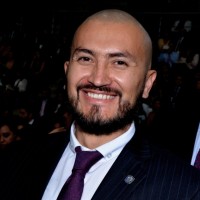}}]{Jhonathan Sora-Cardenas}
received the bachelor's degree in electronics engineering from Universidad Santo Tomás, Bogotá D.C., Colombia, in 2008, and the master's degree in bioengineering from Pontificia Universidad Javeriana, Bogotá D.C., Colombia, in 2018. He is currently pursuing the Ph.D. degree in human-centered computing at the Georgia Institute of Technology, Atlanta, GA, USA.

He is a Graduate Research Assistant affiliated with the People-Agents Research for Computing Education (PARCE) Laboratory and the Culture and Technology (CAT) Laboratory, under the supervision of Dr. Pedro Guillermo Feijóo-García and Dr. Betsy DiSalvo. His research examines human-centered AI and computing education, including students' internship experiences, self-efficacy, the impostor phenomenon, and trust in AI-enabled systems. He previously held teaching positions at Universidad ECCI and Pontificia Universidad Javeriana, both in Bogotá D.C., Colombia.
\end{IEEEbiography}

\vspace{-1em}
\begin{IEEEbiography}[{\includegraphics[width=1in,height=1.5 in,clip,keepaspectratio]{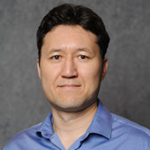}}]{Aibek Musaev} received the B.Sc. degree in computer science in 1999, the M.Sc. degree in computer science in 2000, and the Ph.D. degree in computer science in 2016 from the Georgia Institute of Technology, Atlanta, GA, USA.

He is a Senior Lecturer in the School of Computing Instruction in the College of Computing at the Georgia Institute of Technology, Atlanta, GA, USA. His research spans artificial intelligence and data science, with applications in social computing, crisis informatics, cyber-physical sensing, and computing education. He previously held a faculty position at The University of Alabama, Tuscaloosa, AL, USA. He received the Scholarship of the President of the Kyrgyz Republic in 1996 and the Collaborative Arts Research Initiative (CARI) Faculty Fellowship in 2019.
\end{IEEEbiography}

\vspace{-1em}
\begin{IEEEbiography}[{\includegraphics[width=1in,height=1.5 in,clip,keepaspectratio]{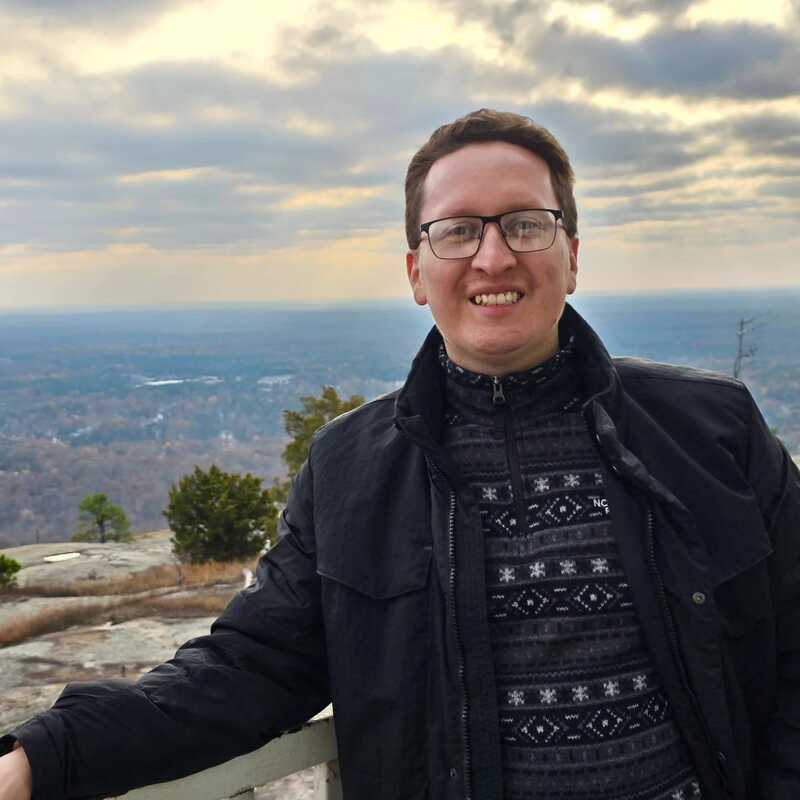}}]{Pedro Guillermo Feijóo-García}
(Member, IEEE) received the B.Sc. degree in systems and computing engineering in 2013, the B.Sc. degree in mechanical engineering in 2014, and the M.Sc. degree in systems and computing engineering in 2015 from Universidad de los Andes, Bogotá D.C., Colombia. He received the Graduate Certificate in University Teaching from Universidad El Bosque, Bogotá D.C., Colombia, in 2018, and the Ph.D. degree in human-centered computing from the University of Florida, Gainesville, FL, USA, in 2023.

He is a Lecturer in the School of Computing Instruction and Director of the People-Agents Research for Computing Education (PARCE) Laboratory in the College of Computing at the Georgia Institute of Technology, Atlanta, GA, USA. His research focuses on human-centered computing, computing education, and human-AI interaction, with particular emphasis on hidden curriculum, software engineering education, and AI-supported learning environments. He is a Fulbright alumnus with more than 15 years of teaching experience in higher education across Colombia and the USA.
\end{IEEEbiography}

\end{document}